\documentclass[aps,pra,twocolumn,groupedaddress]{revtex4}
\usepackage{mathrsfs}
\usepackage{graphicx}
\usepackage{epstopdf}
\usepackage{psfrag} 	
\usepackage{color}
\usepackage{epsfig}
\usepackage{stmaryrd}
\usepackage[utf8x]{inputenc}
\newcommand{\bra}[1]{\langle{#1}|}
\newcommand{\ket}[1]{|{#1}\rangle}

\newcommand{\ev}[1]{\langle{#1}\rangle}       			%
\newcommand{\op}[1]{\mathbf{#1}}              				%

\begin{document}
\title{The Influence of Geometry on the Vibronic Spectra of Quantum Aggregates}

\author{Alexander Eisfeld$^1$}
\email{eisfeld@mpipks-dresden.mpg.de}
\author{Georg Schulz$^{2,3}$}
\email{Georg.Schulz@unibas.ch}
\author{John Briggs$^{1,2}$}
\email{briggs@physik.uni-freiburg.de}

\affiliation{$^1$Max-Planck-Institute for the Physics of Complex Systems, Dresden, Germany,\\$^2$Theoretical Quantum Dynamics, Physics Institute, University of Freiburg, Germany, \\$^3$Biomaterials Science Center (BMC), University of Basel,
 Switzerland }

\date{\today}

\begin{abstract}
A study is presented of the localisation of excitonic states on extended
molecular aggregates composed of \emph{identical} monomers arising, not from
disorder due to statistical energy shifts of the monomers, induced by environmental interactions (Anderson localisation), but rather simply due to changes in the orientation and geometrical arrangement of the transition dipoles. It is shown further that such small changes nevertheless can have a drastic effect on the shape of the vibronic spectrum of the aggregate. The vibronic spectra are calculated using the "coherent exciton scattering" (CES) approximation whose derivation we generalise to be applicable to aggregates of arbitrary size and geometry.

\end{abstract}

\maketitle 
\section{Introduction}
  The quantum behaviour of extended aggregates of atomic and molecular monomers, containing from just a few, up to thousands of sub-units, is attracting increasing attention in chemistry and physics. Prominent examples are aggregates of large dye molecules \cite{Ko96__,KiDa06_20363_,EiKnKi09_658_}, chromophore assemblies describing the photosynthetic unit \cite{AmVaGr00__,ReMaKue01_137_} or assemblies of ultra-cold atoms \cite{WAtEi10_053004_,RoHeTo04_042703_,MueBlAm07_090601_,AtEiRo08_045030_}.  
 Clearly, since the interacting transition dipoles are three-dimensional vectors (the interaction is of tensor form) and since the probing light can be polarised, the precise geometry of the aggregate is of crucial importance. These two effects, the relative orientation of neighbouring dipoles and their orientation with respect to the light polarisation, decide both the energy dependence of the aggregate density of states and the distribution of oscillator strength. These are the key elements in deciding the response of the aggregate to incident light and how the electronic excitation propagates subsequently. In the case of aggregates of large molecules this has been appreciated for a very long time and measured spectra have been used to help infer the usually unknown geometry of the aggregate under observation. For example, whether dye aggregates exhibit a  J-band, an H-band, or both, in  absorption provides strong clues as to aggregate geometry.
 
  Recently two striking new effects of the influence of geometry on spectral absorption have been observed. 
When pressure is applied to cylindrical aggregates of dye molecules, one observes significant spectral shifts \cite{LiCh96_5359_,Sp99__,SpDa98_738_}
which are interpreted \cite{Sp99__,SpDa98_738_}  as arising from a collapse of
the cylinder cross-section from circular to elliptical form.
 Similarly in spectral studies of the light-harvesting complex \cite{NoRuGr06_2890_,ElFr08_386_,MiKiSa08_16759_,RiBaSo07_20280_} it has been shown that departures from a circular arrangement lead to strong spectral changes and detailed comparison with experiment allows one to predict one particular geometrical arrangement of the chromophores as  most likely.
 
 Several methods have been put forward to account for the continuous broadening of observed spectra due to interaction with vibrational and environment degrees of freedom. By far the most popular method  is to ignore intra-monomer vibrations and  the specific nature of external degrees of freedom coupling to the exciton and simply to assume that the monomer transition energies (diagonal disorder) and/or the inter-monomer electronic interaction strengths (off-diagonal disorder) are randomly distributed according to some prescription. In this model, transition energies with random deviations from the mean act effectively as 'impurities' in the aggregate leading to 'Anderson'  localised states which  are instrumental in inhibiting energy transfer, as seen  for example in \cite{DoMa04_226_,BeMaKn03_217401_}.
 
  One of the main points to be emphasised below is that geometrical conformation, even in a model of  monomers with identical transition energies, can lead to similar effects without random disorder in the couplings. That  is, abrupt  local changes in conformation, e.g. simple bending of a polymer chain, give rise to changes in absorption spectra and occurrence of localised states.
This is in contrast to 'Anderson' localisation arising from statistical fluctuations. We feel that insufficient attention has been accorded to the strong spectral effect of such simple 'geometrical'  localisation. Although 'topological' localisation has been well-studied in other contexts, again it has been from a statistical point of view.

 Of course, the formation of localised states, splitting off from delocalised band of states, due to isolated impurities is a well-known phenomenon. It occurs not only in exciton bands but also in conduction bands and phonon bands. However, in our case there is no real impurity, just a change in interaction due to changed geometry.
As illustrative examples of this we will consider the very simplest cases of a single bend in a linear chain of monomers or an elliptical deformation of a circle. 

Many related studies have been presented  concentrating on statistical distributions of "impurities". Dynamical diagonal disorder (i.e.\ time-dependent fluctuations of monomer transition energies) on circular aggregates has been studied by  Bakalis et.~al.~\cite{BaCoKn99_2208_} and by Wubs and Knoester \cite{WuKn98_63_}. Similarly Freiberg et.~al.~\cite{FrRaeTi09_102_} studied disorder in the LH1 and LH2 photosynthetic aggregates  and Barvik et.~al~\cite{BaWaNe99_173_} calculated lineshapes for rings of BChl monomers including statistical disorder in transition energies and coupling strengths but only for dichotomic disorder and for averaged nearest-neighbour coupling. In a further advance, Fidder et.~al.~\cite{FiKnWi91_7880_}, the effect of off-diagonal disorder in coupling strengths was included via a statistical distribution of inter-monomer spacings. Wu and Small \cite{WuSm98_888_} also considered both diagonal and off-diagonal statistical disorder on circular aggregates with reference to the LH2 photosynthetic complex. The question of the influence of elliptical deformation of circular structures encountered in LH2 photosynthesis was considered by Matsushita et.~al.~\cite{MaKeOi01_1604_} for purely electronic transitions and with inclusion of statistical broadening by Warns et.~al.~\cite{WaReBa03_1_}.

 Here we seek to include vibronic coupling more directly than in the statistical models, which essentially ignore internal vibrational structure. Even though one is dealing with large organic molecules, in statistical models they are treated as structureless entities characterised only by a given electronic transition energy.  As in our previous work on spectra and energy transfer \cite{EiBr06_376_,EiKnBr07_104904_,RoScEi09_044909_}, we consider first the purely electronic model of {\it{identical}} coupled monomers and then the realistic case of  broad structured spectra arising from interaction with both intra-molecular vibrational modes and with external vibrational modes of the environment. We use the "coherent exciton scattering" (CES) approximation generalised to be applicable to aggregates of arbitrary size and geometry.
This approach can also take the influence of disorder and and temperature implicitly into account \cite{EiBr06_113003_}. In a previous publication \cite{RoEiBr08_258_} we have tested the CES approximation by diagonalising Hamiltonians of the type considered here. For red-shifted (J-band) aggregate spectra the CES calculation gives very good agreement with exact results for all coupling strengths between monomers. For blue shifts (H-band spectra) it gives good agreement for weak and strong coupling. Only in the intermediate coupling case is the agreement not so good, although the main qualitative features of the full diagonalised results are still obtained.
This theoretical study has been performed for the extreme case of a single vibrational mode. 
Comparison to experiment (where the vibrations form a continuum)  has shown that in such realistic situations the CES approximation is able to reproduce the J-band \cite{EiBr07_354_} and also the H-band \cite{EiBr06_376_} in the intermediate coupling regime with remarkable accuracy. 

 Without the influence of vibrations, the overall spectral structure is decided  by the density and oscillator strength distribution of electronic transitions. Hence, we will show how the geometry of transition dipole orientation and  isolated geometrical irregularities can seriously influence the purely electronic spectrum of  molecular aggregates.  In particular we point out that the orientation of transition dipoles plays a decisive role in deciding the characteristics of absorption spectra. Also, depending on the orientation, sudden changes in conformation due to bends or kinks in regular structures lead to local changes in off-diagonal coupling and thus to localisation of wave-functions on these sites.  In some cases, these localised states can still be prominent in the spectrum when vibrational structure is taken into account explicitly. 

  The plan of the paper is as follows. In section II  we derive expressions for the aggregate absorption spectrum for the purely electronic case and when coupling to vibrations are included. The simplest case of a bent linear chain is considered in section III and that of a circle deformed into an ellipse in section IV before conclusions are discussed in section V.
 
\section{The aggregate spectrum}
\subsection{the purely electronic case}

 The purely electronic model in which the inter-monomer interaction is taken to be that of point dipoles located on each of $N$ monomers is particularly simple. The total aggregate Hamiltonian is $\op{H}=\op{H}_0+\op{V}$, where $\op{H_0}$ is the sum of $N$ identical monomer electronic Hamiltonians and $\op{V}$ is the dipole-dipole coupling operator. First the Hamiltonian is expressed in the localised basis of aggregate states,
\begin{equation}
	\ket{\pi_n}:=\ket{\phi_n^e}\prod_{m\neq n}\ket{\phi_m^g}
	\label{equ:pi_n}
\end{equation}
where $\ket{\phi_n^e}$ is the excited electronic state and $\ket{\phi_n^g}$ the ground electronic state of monomer $n$. In this basis the Hamiltonian reads,

\begin{equation}
\op{H} = \sum_n\epsilon_n \ket{\pi_n}\bra{\pi_n} + \sum_{n,m}V_{nm} \ket{\pi_n}\bra{\pi_m} ,
\end{equation}
where $\epsilon_n$ is the transition energy of monomer $n$ and the aggregate ground state energy is taken to be zero. This is the starting point for statistical models in which $\epsilon_n$ and $V_{nm}$ are considered randomly distributed to include  vibrational and other  broadening phenomenologically. Here, starting first with the purely electronic problem for identical monomers,  we take all $\epsilon_n$ to be the same and the $V_{nm}$ to be determined from the dipole-dipole interaction
\begin{equation}
\label{eq:Vnm}
V_{nm} =  \frac{\vec{\mu}_n.\vec{\mu}_m}{|X_{nm}|^3} - 3 \frac{\vec{\mu}_n.\vec{X}_{nm} \quad\vec{\mu}_m.\vec{X}_{nm} }{|X_{nm}|^5}
\end{equation}
where $\vec{\mu}_n$ is the transition dipole on monomer $n$ and $\vec{X}_{nm}$ is the vector separation of monomers $n$ and $m$.\\
Below the absorption spectrum is expressed in terms of the energy-dependent Green function defined by the aggregate Hamiltonian as,
\begin{equation}
\op{G}(E)=(E-\op{H}+i\delta)^{-1},\qquad\delta=0_+
	\label{equ:energie_greens_agg}
\end{equation}
Similarly, for the non-interacting monomers we have,
\begin{equation}
	\op{g}(E)=(E-\op{H}_0+i\delta)^{-1},
	\label{equ:energie_greens_mon}
\end{equation}
The two Green operators are connected by the equation,
\begin{equation}
	\op{G}(E) = \op{g}(E) +\op{g}(E)\op{V}\op{G}(E).
	\label{equ:dyson}
\end{equation}
In the localised electronic basis this equation reads
\begin{equation}
	G_{nm}=g_n\delta_{nm}+g_n\sum_{n^{\prime}}V_{nn^{\prime}}G_{n^{\prime}m}.
	\label{equ:dyson_gleichung}
\end{equation}
As alternative,which as we will see exhibits geometric effects more clearly, we transform to the delocalised exciton basis,
\begin{equation}
\ket{k} = \sum_{n}a_{kn}\ket{\pi_n}, \qquad k = 1\dots N
\label{equ:transformation}		
\end{equation}
where the coefficients $a_{kn}$ are determined by diagonalising the aggregate Hamiltonian $\op{H}$.  Note that in this paper we use exclusively the subscripts $n,m$ to denote the "site" basis and the subscripts $k,k^\prime$ to denote the exciton basis.
In the exciton basis one readily sees that the Hamiltonian has matrix element 
\begin{equation}
\bra{k} \op{H}\ket{k^\prime} =  (\epsilon  + C_k	) \delta_{kk^\prime},
\end{equation}
where,
\begin{equation}
\label{eq:C_k}
C_k = \sum_{nm} V_{nm} a_{nk} a_{mk}^{*}	
\label{equ:coupling}
\end{equation}

In the purely electronic, one-level model one has the simple form $g_n(E) = (E - \epsilon_n + i \delta_+)^{-1}$.
Hence, for identical monomers, Eq.(\ref{equ:dyson_gleichung}) simplifies since $g_n = g$ is independent of $n$, i.e.,
\begin{equation}
	G_{nm}=g\delta_{nm}+g\sum_{n^{\prime}}V_{nn^{\prime}}G_{n^{\prime}m}.
	\label{equ:dyson_gleichung2}
\end{equation}
When transformed to the exciton basis this equation reads,
\begin{equation}
G_{kk^\prime} = g\delta_{kk^\prime} + gC_k G_{kk^\prime}.
\label{equ:dyson_gleichungk}
\end{equation}
or,
\begin{equation}
G_{kk^\prime} = \frac{ g }{(1 - gC_k )}\delta_{kk^\prime}
\label{equ:dysonfrac}.
\end{equation}
Since $g(E) = (E - \epsilon + i \delta_+)^{-1}$, from the above equation (\ref{equ:dysonfrac}) one has 
\begin{equation}
G_{kk^\prime}(E) =  \frac{1}{(E - \epsilon - C_k + i \delta_+)} \delta_{kk^\prime},
\end{equation}
 as must be since $\op{H}$,  and therefore $\op{G}$, is diagonal in the exciton basis with eigenergies $\epsilon + C_k$. Although trivial in this case, the above formula (\ref{equ:dysonfrac}) will be useful when vibrations are included.
The shape of the absorption spectrum is given by $- {\rm Im} A(E)$ where the spectral function $A(E)$ is \cite{BrHe71_865_,BrHe70_1663_,EiKnBr07_104904_},
\begin{equation}
A(E) = \sum_{nm} (\vec{e}.\vec{\mu}_n)(\vec{\mu}_m.\vec{e}^{*}) G_{nm}(E)
\label{equ:absorption_nm}
\end{equation}
where $\vec{\mu}_n$ is the transition dipole on monomer $n$and $ \vec{e}$ is the light polarization vector.
Transformed to the diagonal exciton representation, this equation becomes,
\begin{equation}
A(E) = \sum_{k}\left| (\vec{e}.\vec{\mu}_k)\right|^2  G_{kk}(E) =  \sum_{k}\left| (\vec{e}.\vec{\mu}_k) \right|^2\frac{ g(E) }{(1 - g(E) C_k )},
\label{equ:absorption_kk}
\end{equation}
where the effective transition dipole in exciton state $k$ is given by $\vec{\mu}_k = \sum_{n}\vec{\mu}_n a_{nj}$. This equation makes clear how the geometry influences the absorption spectrum via both $\vec{\mu}_k$ and in the interaction $C_k$. Clearly, if the intra-monomer interaction is switched off, then the monomer spectral function is
\begin{equation}
M(E) = \big{(} \sum_{n}\left| (\vec{e}.\vec{\mu}_n)\right|^2\big{)} g(E)
\label{equ:absorptionmonn}
\end{equation}
or, equivalently, in the exciton basis,
\begin{equation}
M(E) = \big{(}\sum_{k}\left| (\vec{e}.\vec{\mu}_k)\right|^2\big{)} g(E)
\label{equ:absorptionmonk}
\end{equation}\\

\subsection{Inclusion of Vibronic Coupling}

\begin{figure*}[hpt]
\psfrag{state}{}
\psfrag{number}{}
\psfrag{EnergInV}{}
\psfrag{Wavenumber}{}
\psfrag{Energy}{\scriptsize Energy}
\psfrag{AbsStr}{\scriptsize absorption}
\psfrag{Abs}{\scriptsize absorption}

\psfrag{2.33338}{\bf a)}
\psfrag{2.77268}{\bf b)}
\psfrag{3.58581}{\bf c)}
 \includegraphics[width=1.5\columnwidth]{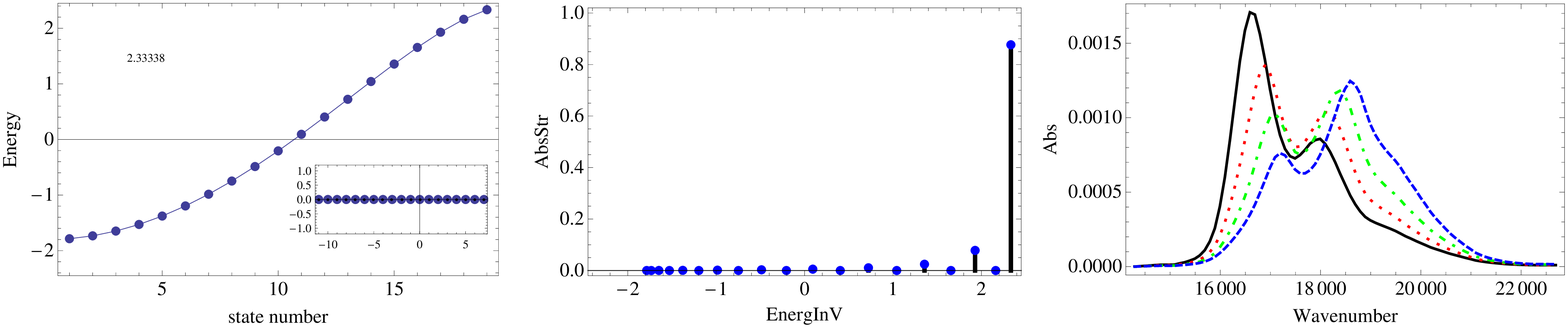}%
\\
\vspace{-0.35cm}
\includegraphics[width=1.5\columnwidth]{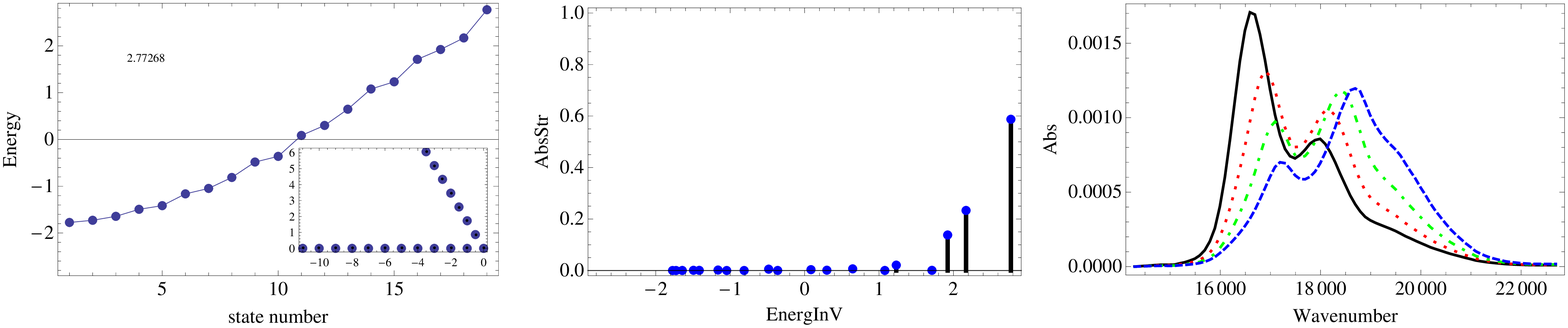}%
\\
\vspace{-0.35cm}
\psfrag{state}{\scriptsize state number }
\psfrag{EnergInV}{energy}
\psfrag{Wavenumber}{\scriptsize energy in cm$^{-1}$}
\includegraphics[width=1.5\columnwidth]{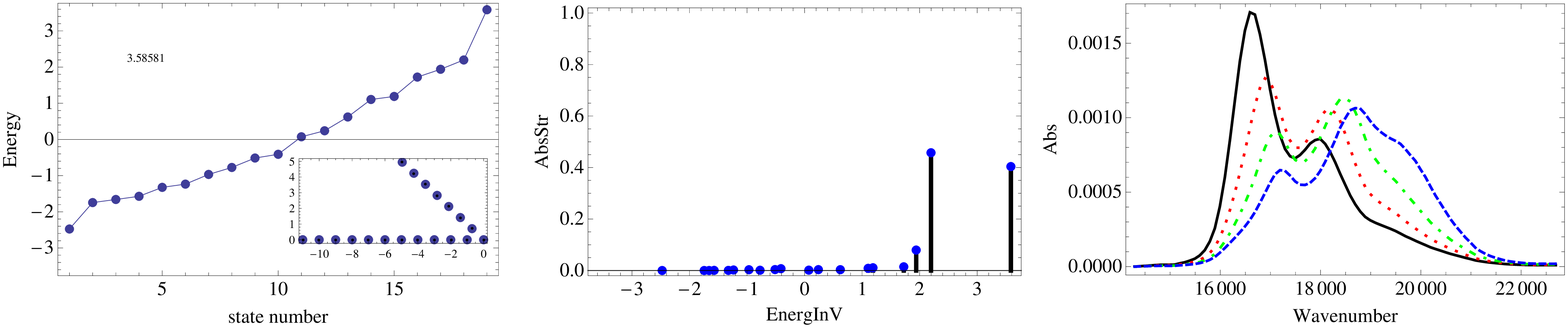}%
\caption{\label{fig:knick_perp}The energy eigenvalues (left column) and electronic spectra (middle column) for a linear chain of monomers with $\hat\mu = (0,0,1)$ and bend angle 
$\Phi=0^\circ,120^\circ,135^\circ$ in a), b), c) respectively. The vibronic CES spectra (right column) are calculated with $V=150$ cm$^{-1}$ (red dotted), $V=300$ cm$^{-1}$ (green dash-dotted) and $V=450$ cm$^{-1}$ (blue dashed).   }
\end{figure*}

To calculate realistic spectra one must include the broadening due to coupling to both IM and EM vibrational degrees of freedom. This requires inclusion of these degrees of freedom in the definition of the non-dipole coupled monomer Hamiltonian $\op{H}_0$. When this is done, the Green operators $\op{g(E)}$ and $\op{G(E)}$ become dependent on these degrees of freedom. Then the spectral function (\ref{equ:absorption_nm}) must be generalised to \cite{BrHe70_1663_,EiBr02_61_,EiKnBr07_104904_}
\begin{equation}
A(E) = \sum_{nm} (\vec{e}.\vec{\mu}_n)(\vec{\mu}_m.\vec{e}^{*}) \ev{G_{nm}(E)},
\label{equ:absorptionvib}
\end{equation}
where $\big{<}....\big{>}$ denotes an integration over the full vibrational ground state of the aggregate.
Note however, that Eq.(\ref{equ:absorption_kk}) cannot be so simply generalised since $G$ is no longer diagonal in the pure exciton basis. Nevertheless, we can still transform to this basis, when
Eq.(\ref{equ:absorption_kk}) is replaced by,
\begin{equation}
A(E) = \sum_{kk^\prime} (\vec{e}.\vec{\mu}_k^\prime).(\vec{\mu}_k.\vec{e}^{*}) \ev{G_{kk^\prime}(E)}
\end{equation}
Similarly in this basis, Eq.(\ref{equ:dyson}) becomes
\begin{equation}
\ev{G_{kk^\prime}} = \ev{g_{kk^\prime}} + \ev{(\op{gVG})_{kk^\prime}}
\label{equ:dysonjk}
\end{equation}
In general this equation cannot be simplified further. To proceed, two strategies are possible. One is to work fully numerically and simply expand the Hilbert space to a basis which is not purely electronic but which includes all vibronic degrees of freedom. Then the Hamiltonian and $\op{G(E)}$ again are diagonalised in this expanded basis. In view of the explosion in the number of vibronic basis states however, this strategy is only possible for very small aggregates and only a few vibrational modes \cite{RoEiDv11_054907_,RoEiBr08_258_,SpZhMe04_10594_,ZhSp05_114701_,ScFi84_269_,BoTrBa99_1633_,AnPe07_155_,RoStEi11_034902_}. Alternatively one can seek approximations. In previous work we have shown that the 'coherent exciton scattering' (CES) approximation gives a good reproduction of experimental aggregate spectra \cite{EiBr02_61_,EiBr06_376_,EiBr07_354_,EiKnBr07_104904_} and agrees with  exact diagonalisation  results for a variety of examples \cite{RoEiBr08_258_,EiBr06_113003_}. In this approximation, the operator $\op{g}$ in the second term on the right hand side of Eq.~(\ref{equ:dysonjk}) is replaced by its ground-state vibrational average $\ev{\op{g}}$. Then, if again we assume that $\ev{g_n}$ is independent of $n$ due to identity of the monomers, it is easy to show that Eq.~(\ref{equ:dysonjk}) reduces to,
\begin{equation}
\ev{G_{kk^\prime}} = \ev{g} \delta_{kk^\prime} + \ev{g} C_k\ev{G_{kk^\prime}}
\label{equ:dysonkkAv}
\end{equation}
with $C_k$ given by Eq.~(\ref{equ:coupling}).
Then one has again the simple result,
\begin{equation}
\ev{G_{kk^\prime}} = \frac{ \ev{g} }{(1 - \ev{g}C_k )}\delta_{kk^\prime}
\label{equ:dysonfracAv}.
\end{equation}
and 
\begin{equation}
\label{eq:CESabsorption}
A(E) =  \sum_{k}\left| (\vec{e}.\vec{\mu}_k) \right|^2\frac{ \ev{g(E)} }{(1 - \ev{g(E)}C_k )},
\label{equ:absorption_kkAv}
\end{equation}
which are identical to the purely electronic case Eq.(\ref{equ:dysonfrac}) and Eq.~(\ref{equ:absorption_kk}) except that $g$ is replaced by its vibronic counterpart $\ev{g}$. Similarly, the spectrum of non-interacting monomers is given by Eq.~(\ref{equ:absorptionmonn}) with $g$ replaced by $\ev{g}$. This has the important consequence that all monomer and aggregate spectra are broadened due to simultaneous vibrational excitation and environment coupling. 
The input to the CES spectral calculation is the complex function $\ev{g(E)}$ which we will take to be of such a form as to reproduce typical experimental monomer spectra composed of several, continuously-broadened peaks as shown in Fig.~(\ref{fig:knick_perp}) right column, black curve.

\begin{figure}[bp]
\psfrag{k}{\small eigenstates}
\psfrag{Monomer}[c]{\small Monomer number $m$}
\includegraphics[width=0.9\columnwidth]{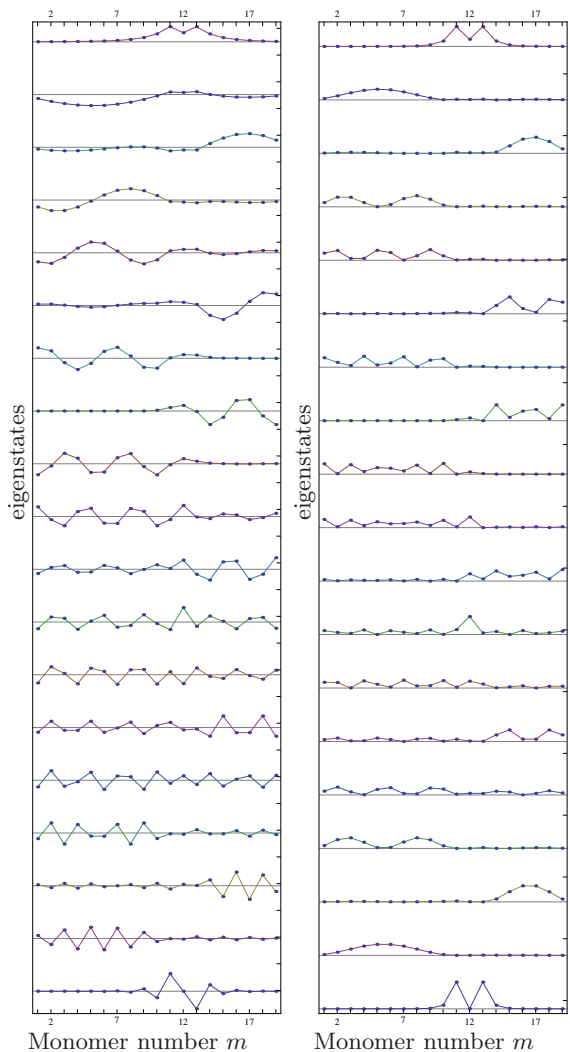}%
\caption{\label{fig:Wavefunct}Wavefunctions, left column, and modulus squared, right column, of a bent linear chain with $\phi=135^{\circ}$ as a function of monomer number. The energy of the states is increasing from bottom to top.
}
\end{figure}

\section{The bent linear chain}

First of all we consider  a rigid linear chain of  equally-spaced identical monomers. A bend  is introduced at one  monomer (we take number 12) to give a V-shaped chain in which the two half-chains are at a mutual angle $\Phi$ (see the insets on Fig.(\ref{fig:knick_perp}), where we have taken $N=19$ monomers and located the bend on monomer 12). Sequences of such bent chain segments are a good model for certain types of conjugated polymers, as studied  for example in Ref.~\cite{CoSc09_369_}.
Although we consider interactions between all monomers on the chain, it is clear that the interactions between the monomer at the vertex and its nearest neighbours will suffer the largest change upon bending. Indeed we will find that this simple symmetry breaking causes the appearance of localised states centred around the vertex.
The linear case $\Phi = 0$ is shown in Fig.~(\ref{fig:knick_perp}a). Here the dipoles have been taken to be aligned perpendicular to the chain. The nearest-neighbour interaction from Eq.~(\ref{eq:Vnm}) for this geometry will be taken as the unit of energy in all cases. Restricting to nearest-neighbours alone, in this unit the exciton band would occupy the domain $-2$ to $+2$ . Including all couplings shifts the band slightly, as seen in Fig.~(\ref{fig:knick_perp}a) and discussed e.g.\ in \cite{FiKnWi91_7880_}.

 For each geometry we calculate the excitonic eigenvalues and eigenvectors (wave-functions) by direct numerical diagonalisation and in the first and second columns of  Fig.~(\ref{fig:knick_perp})
 we present the energy eigenvalues and oscillator strength 'stick' spectra respectively. Here and in subsequent figures showing spectra, the undistorted nearest-neighbour value of $V_{n,n+1}$ in Eq.~(\ref{eq:C_k}), which we call the coupling strength $V$, is taken as the energy unit for the stick spectra in the second column. Note also that we have normalised the oscillator strengths to sum to unity in each case. Only in assembling the CES vibronic spectrum shown in the third column do we need absolute values. Here we have taken as typical monomer spectrum that of pinacyanol, stretching from roughly 17000  to 20000 cm$^{-1}$  and calculated the aggregate spectra using Eq.~(\ref{eq:CESabsorption}). Then the couplings $C_k$ must also be absolute, which is achieved by fixing $V$.
The realistic size of this is determined by the width $\Delta \approx 2000$ cm$^{-1}$ of the monomer spectrum. In Fig.~(\ref{fig:knick_perp}) and in subsequent figures we plot the continuous vibronic spectra in absolute energy units of $cm^{-1}$.

For the straight chain, shown in the top row of Fig.~\ref{fig:knick_perp}a, the results are well-known. The energy spectrum is a smooth function of $k$ and one level at the band edge (a blue-shifted so-called H band in this geometry) carries almost all the oscillator strength. Inspection of the wave-functions (not shown) shows that all levels are delocalised over the whole chain. In the right column we show the CES spectrum for the monomer and for weak, ($V = 150$ cm$^{-1}$), intermediate ($V = 300$ cm$^{-1}$) and strong ($V = 450$ cm$^{-1}$) coupling.
As the coupling strength increases, the monomer spectrum metamorphoses into a typical, relatively broad blue-shifted H band.
Note, that the distinction weak/intermediate/strong is not based on the Simpson-Peterson criterion\cite{SiPe57_588_} but is just introduced to denote the relative strengths of the couplings.
\begin{figure*}[tp]
\psfrag{4.00751}{\bf a)}
\psfrag{4.74309}{\bf b)}
\psfrag{state}{}
\psfrag{number}{}
\psfrag{EnergInV}{}
\psfrag{Wavenumber}{}
\includegraphics[width=1.5\columnwidth]{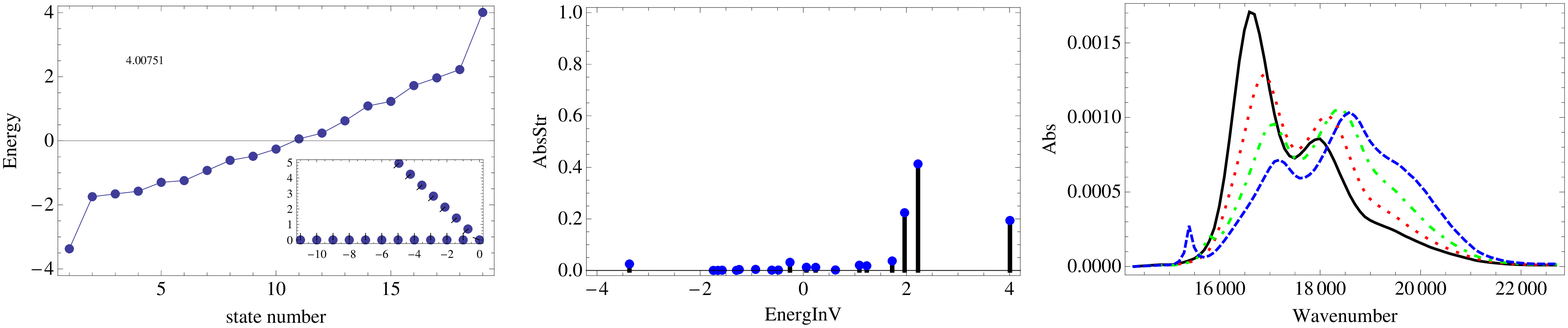}%
\\
\psfrag{state}{\scriptsize state number }
\psfrag{EnergInV}{energy}
\psfrag{Wavenumber}{\scriptsize energy in cm$^{-1}$}
\includegraphics[width=1.5\columnwidth]{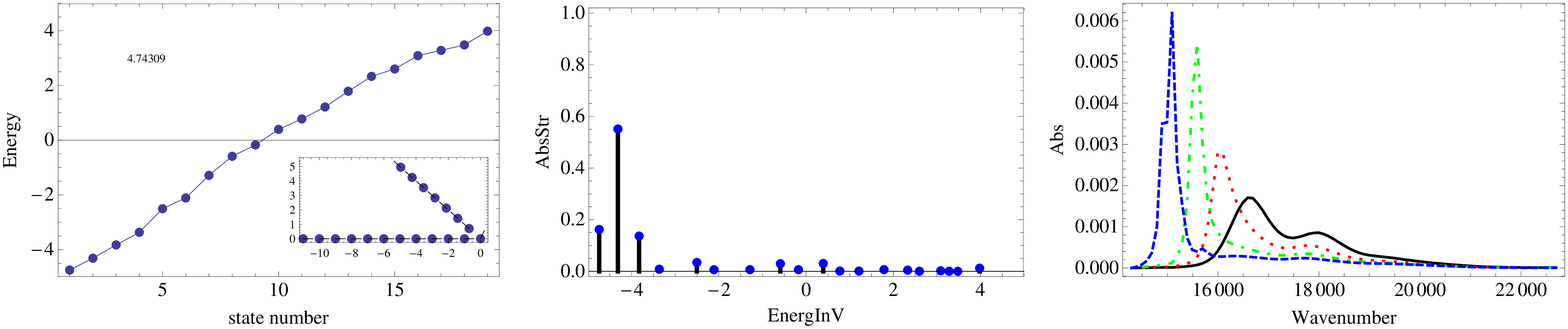}%
\caption{\label{fig:knick_para} Eigenvalues and spectra as in Fig.(1) for  a bent chain with $\phi=135^\circ$ and dipoles with orientation a) $\hat{\mu}=(0,1,1)/\sqrt{2}$ and  b) $\hat{\mu}=(1,0,0)$. In the right column the CES spectra are for the same values $V$ of the coupling as in Fig.(1).  }
\end{figure*}

We take the bent chain to lie in the $(x,y)$ plane. Calculations for finite $\Phi$ show little change both in the pure electronic and the vibronic spectrum so long as $\Phi \le 90^{\circ}.$ However, for obtuse angles there begin changes in the spectrum, as shown in Fig.(\ref{fig:knick_perp}b) for $\Phi = 120^{\circ}.$ Here the highest level carrying most oscillator strength has begun to separate from the rest of the exciton band levels. This is seen very clearly in Fig.(\ref{fig:knick_perp}c) for $\Phi = 135^{\circ}.$ Now two levels, one above and one below the band, have split off . An inspection of the 
wave-functions (Fig.~\ref{fig:Wavefunct}) allows a simple explanation in that the two levels which split off are strongly localised on only two monomers. This is our first example of 'geometrical' localisation. The sharp bend in the chain gives a strong interaction between the two monomers adjacent to the vertex monomer. This leads effectively to the formation of an isolated dimer 'impurity' between these two monomers. The pair of states split off from the band in Fig.(\ref{fig:knick_perp}c) are just the symmetric $'+'$ and antisymmetric $'-'$ states of this dimer whose wave-functions are localised on just these two monomers, see the lowest and highest energy states of Fig.\ref{fig:Wavefunct}. For this dipole orientation, only the high-energy symmetric state carries oscillator strength, as seen in Fig.(\ref{fig:knick_perp}c). The actual vertex monomer has now become effectively isolated from the rest of the chain and leads to a further state localised on just this monomer. However this state sits in the middle of the band (state number 8) and carries little oscillator strength. All other states are delocalised but, since the chain now consists of two halves, they are predominantly confined to one or other half of the chain, as shown clearly in Fig.(2). Apart from the highest energy dimer $'+'$ state the delocalised in-phase second and third states from the top of the band are the only ones to carry oscillator strength, as seen in the middle column of Fig.(\ref{fig:knick_perp}b). Inclusion of vibrations, right column of  Fig.(\ref{fig:knick_perp}c) shows the expected formation of an H band
for weak and intermediate coupling, but for strong coupling the isolated,
 dimer peak gives rise to a more pronounced shoulder
 on the spectrum.
Note however, that due to the inclusion of vibrations the highest electronic absorption peak is now accompanied by a complicated vibrational structure. This in turn makes it difficult to distinguish electronic and vibrational contibutions from measured spectra, similar to the situation in dimers \cite{Ei07_321_,EiSeEn08_186_}.

To illustrate the  spectral changes resulting from relatively simple changes
in dipole orientation we now consider a bend in which the dipoles are still
perpendicular to the chain axis but now lie in the plane of the V-shaped bent
chain, not perpendicular to it as above. The first case we consider is where the dipoles have $\mu$ perpendicular to the chain but with equal components in- and out of-plane, shown in Fig.~\ref{fig:knick_para}a for $\Phi = 135^\circ$. Of course the straight chain spectrum 
is again that of Fig.~\ref{fig:knick_perp}a and as the chain is bent, up to $\Phi \approx
120^{\circ}$, the situation is roughly as in Fig.(\ref{fig:knick_perp}b) with the formation of a broad
H band for strong coupling. These absorbing states are delocalised. However,
for $\Phi = 135^\circ$  the split-off dimer localised states again become
evident above and below the band,
but now both symmetric and antisymmetric localised states carry oscillator strength. Then as shown in Fig.~(\ref{fig:knick_perp}), for $\Phi = 135^{\circ}$ sharp peaks appear   both above and below the exciton band. In fact the sharp peak below the band (appearing for strong coupling) is reminiscent of a collective fully delocalised J-band exciton state but we emphasise that it is of completely different character in that its wave-function is localised on just the two monomers adjacent to the vertex.

This similarity of spectral appearance of localised and delocalised states is also seen when we consider a linear aggregate showing a J-band and not an H-band as in Figs.~(\ref{fig:knick_perp}). This is the case when the monomer transition dipoles are all aligned initially parallel to the straight chain as shown in Fig.~(\ref{fig:knick_para}b). The electronic spectrum is dominated by the lowest energy level where all dipoles are in phase. In the vibronic spectrum of the straight chain (not shown) a clear J-band based on this state forms as the coupling strength increases.  Again, for bend angles up to $\approx 120^0$ we find similar spectra to zero degrees i.e.\ this single peak dominates. However, at larger angles, (the case $\Phi=135^{\circ}$ is shown in Fig.~(\ref{fig:knick_para}b) as 
example), again localized dimer states appear above and below the band but now the absorbing $'+'$ state is located below the band. Also, since the strong vertex interaction effectively splits the chain into two linear segments, examination of the wave-function shows that the J band is due to two collective states (the second and third states in the stick spectrum of Fig.~(\ref{fig:knick_para}b) again each largely confined to a separate half of the chain. This gives a  J-band in strong coupling.  On the low energy side of these two peaks appears the localised symmetric dimer state. Note that in the vibronic spectrum of Fig.~(\ref{fig:knick_para}b) these features are only evident when the coupling is strong enough to move their energies below the region of monomer absorption, when the localized state appears as a low-energy shoulder on the J-band.\\
The message of the preceding section is that even in the very simple case of a single fold in a linear chain new spectral features appear which depend crucially on dipole orientation and which make interpretation of spectral features in terms of localised and delocalised states difficult. This will be more so in the case of the complicated geometries occurring in practical cases in biological aggregates for example.

\begin{figure}[bp]
\psfrag{state}{}
\psfrag{number}{}
\psfrag{EnergInV}{}
\psfrag{Energy}{\scriptsize Energy}
\psfrag{AbsStr}{\scriptsize absorption}
\psfrag{4.70636}{a) $\phi=0^{\circ}$}
\includegraphics[width=8cm]{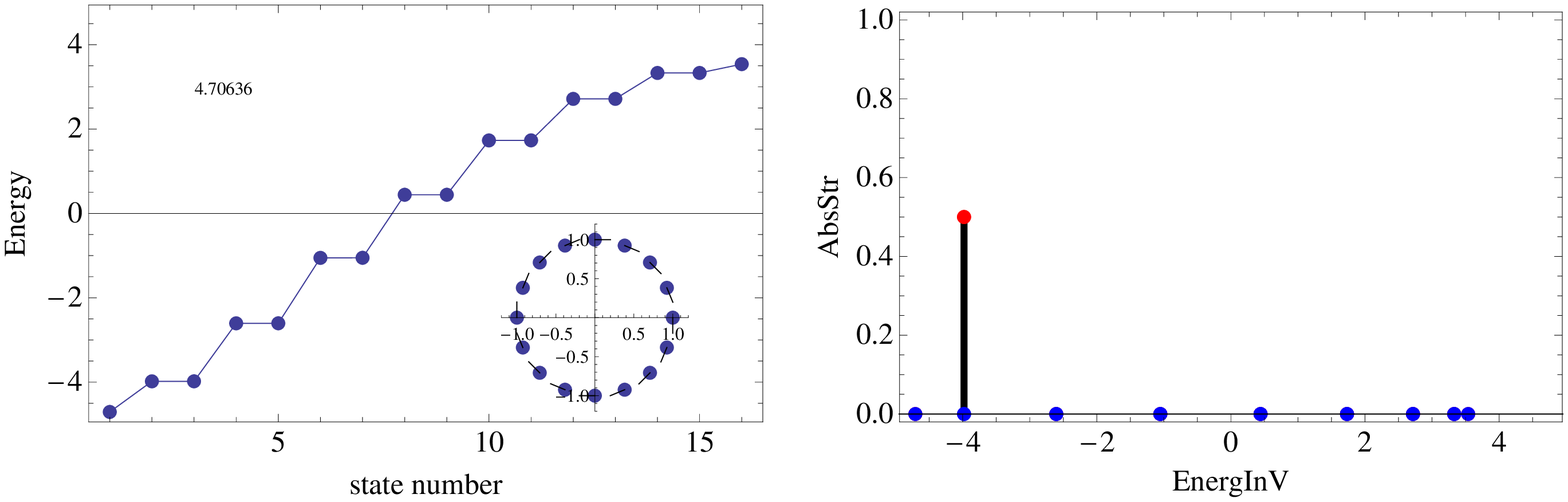}%
\vspace{-0.35cm}\\
\psfrag{0.626366}{b) $\phi=48^{\circ}$}
\includegraphics[width=8cm]{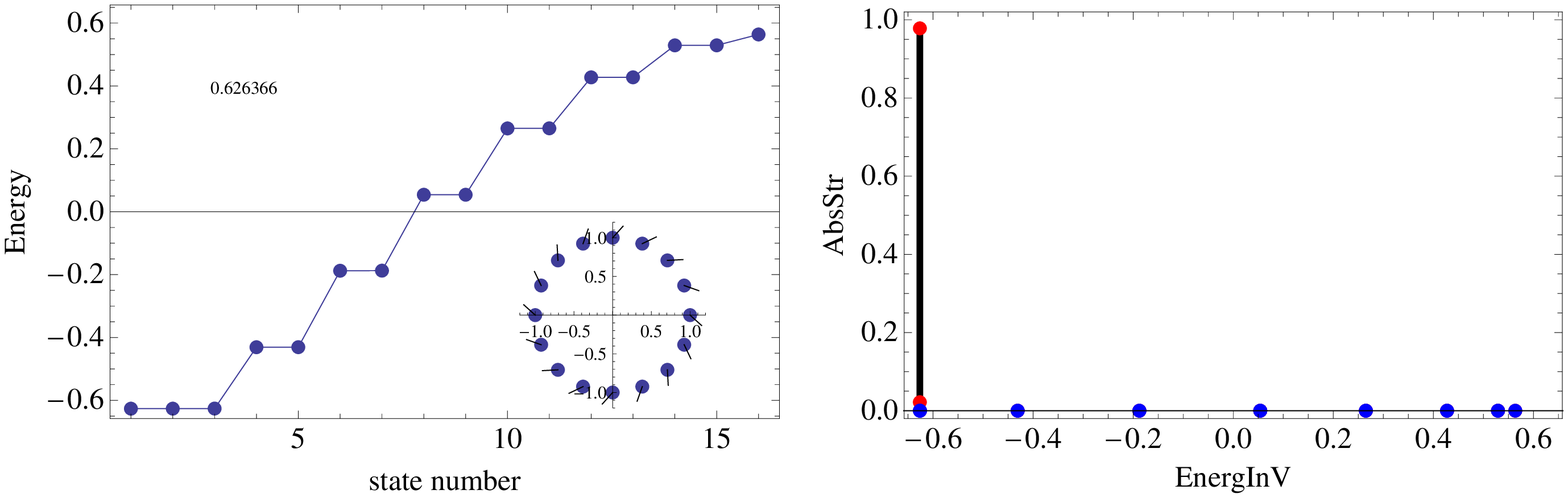}%
\vspace{-0.35cm}\\
\psfrag{0.106787}{c) $\phi=53^{\circ}$}
\includegraphics[width=8cm]{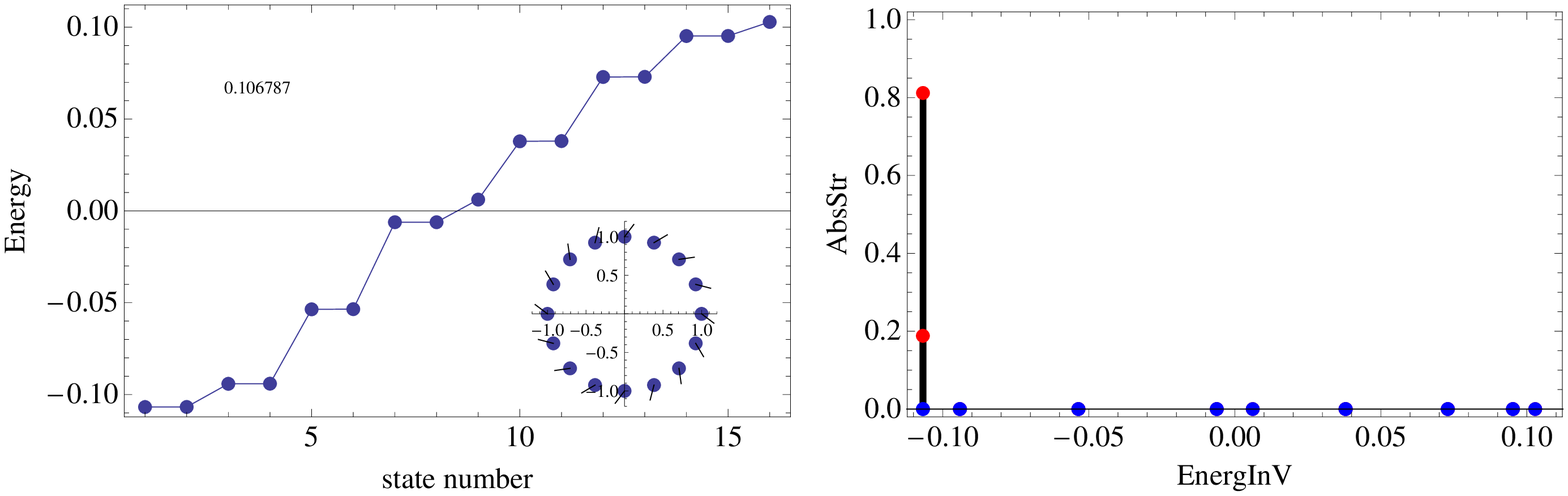}%
\vspace{-0.35cm}\\
\psfrag{0.129454}{d) $\phi=54^{\circ}$}
\includegraphics[width=8cm]{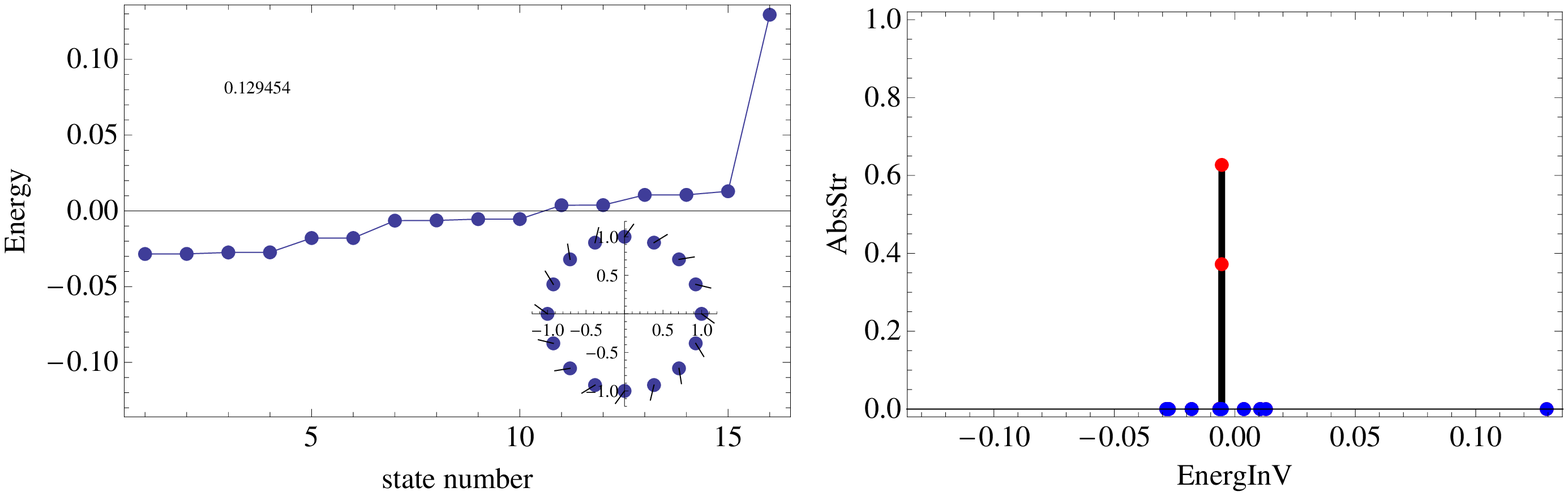}%
\vspace{-0.35cm}\\
\psfrag{0.251375}{e) $\phi=55^{\circ}$}
\includegraphics[width=8cm]{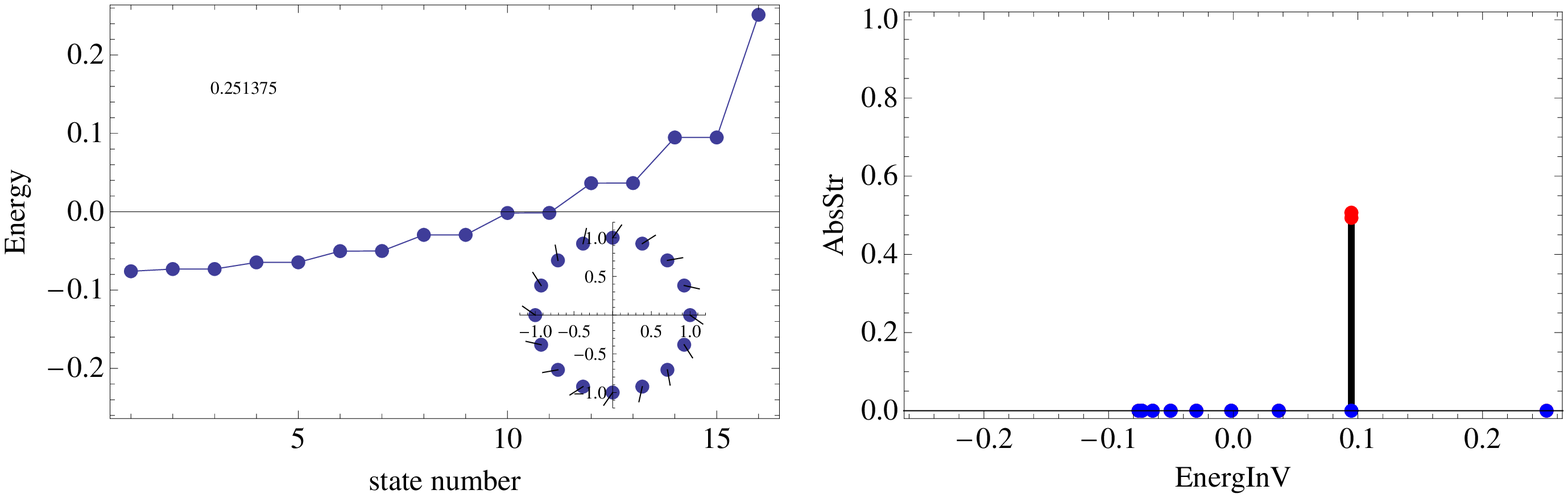}%
\vspace{-0.35cm}\\
\psfrag{2.68211}{f) $\phi=90^{\circ}$}
\psfrag{state}{\scriptsize state number }
\includegraphics[width=8cm]{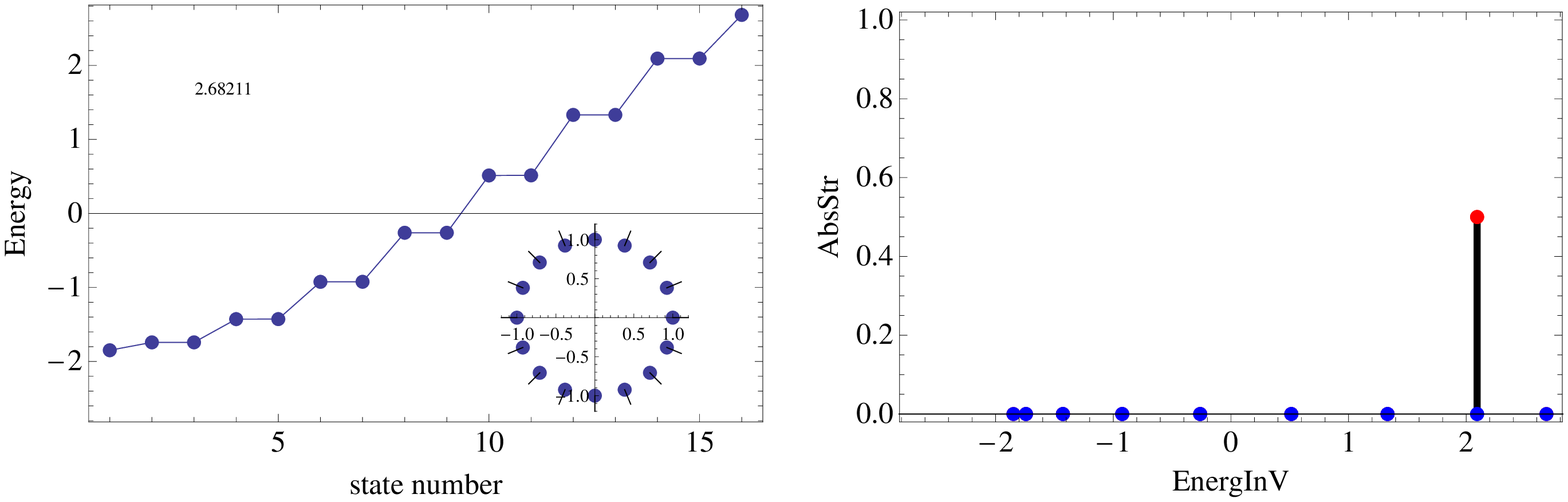}%
\caption{\label{fig:Ring_RadTang} Eigenvalues and stick spectra for a circular aggregate with dipoles lying in the plane at angle $\phi$ w.r.t.\ the tangent, 
a) $\phi=0^{\circ}$
b) $\phi=48^{\circ}$
c) $\phi=53^{\circ}$
d) $\phi=54^{\circ}$
e) $\phi=55^{\circ}$
f) $\phi=90^{\circ}$.  
}
\end{figure}

\section{Circular and Elliptical geometries}
Here we will emphasise the effects of dipole orientation. The first aim is to show that even in completely symmetric geometries, a changing dipole orientation can affect the monomer interactions dramatically and lead to corresponding dramatic changes in the energy location of eigenstates and their light absorbing and emitting properties. This also will show that in general it is absolutely necessary to take all interactions into account: in some cases the much-used nearest-neighbour approximation fails completely. 

The simplest non-linear geometry (but which is practically important in photosynthesis) is a circular arrangement of dipoles. As additional simplification, first we consider equal spacing of dipoles all lying in the plane of the circle. Then we examine the energy and absorption spectrum as a function of the angle $\phi$ of the dipole w.r.t.\ the tangent to the circle i.e.\ $\phi=0^\circ$ is tangential and $\phi=90^\circ$ is radial. The results are shown in Fig.~(\ref{fig:Ring_RadTang}). Each pair of figures is for a different $\phi$ orientation and shows in the left column the geometry and the exciton energy) level distribution (w.r.t.\ $\epsilon = 0$) between the lower and upper band edges. In the right column is shown the corresponding absorption stick spectrum. 
The results can be understood analytically from the  solution using nearest-neighbour coupling only. In this geometry the exciton states are all delocalised over the whole ring and  the eigenenergies are (for $N$ even)
\begin{equation}
E_k = 2 \sum_{n=2}^{N/2} \cos (k n) V_{n,1}+(-1)^j V_{N/2+1,1},
\end{equation}
\begin{equation}
{\rm{with}} \qquad k = \frac{2\pi}{N}j, \qquad j = 0,1......N-1.
\end{equation}

As seen from Fig.(\ref{fig:Ring_RadTang}a) for $\phi=0^\circ$, the levels are doubly degenerate apart from the extreme $j=0$ (all monomers in phase) and $j = N/2$ (successive monomers out of phase) levels which lie at the bottom and top of  the band respectively. Only the two degenerate levels $j=1$ and $j=N-1$ carry oscillator strength. In particular the lowest $j=0$ level in Fig.(\ref{fig:Ring_RadTang}a) is forbidden. Clearly the presence of this dark state below the strongly-absorbing J-band states is important for the fluorescence properties of aggregates with this geometry.  For $\phi=90^\circ$ the energy level ordering is simply reversed and the absorbing states form an H-band with the dark $j=0$ state lying above them in energy.

As $\phi$ increases from zero to approach the 'magic' angle of $\sim 54^\circ$ around which nearest-neighbour coupling vanishes, the major spectral change is that the lowest non-degenerate $j=0$ dark level moves up in energy. In Fig.(\ref{fig:Ring_RadTang}b, first column) for $\phi=48^\circ$ it is state number $3$ and for Fig.(\ref{fig:Ring_RadTang}c)  $\phi=53^\circ$ one sees this non-degenerate level as state number $9$. Note that in this region the nearest-neighbour interaction goes through zero and the higher-order interactions take over. By $\phi=54^\circ$ the $j=0$ level has reached the top of the band (Fig.(\ref{fig:Ring_RadTang}d)) and is being followed by the allowed levels which are now located in the centre of the extremely narrow band.   This band reversal is extraordinarily sensitive to orientation and already by $\phi=55^\circ$  (Fig.(\ref{fig:Ring_RadTang}e))  the H-band configuration has been reached and there is little qualitative change up to $\phi=90^\circ$ of Fig.(\ref{fig:Ring_RadTang}f).  We do not show the broadened CES spectra but it is clear that they change from sharp, fully delocalised J-band character for tangential orientation, through almost unchanged monomer character around $54^\circ$ orientation to H-band character for radially oriented dipoles.

\begin{figure*}[tp]
\psfrag{state}{}
\psfrag{number}{}
\psfrag{EnergInV}{}
\psfrag{Wavenumber}{}
\psfrag{Energy}{\scriptsize Energy}
\psfrag{AbsStr}{\scriptsize absorption}
\psfrag{Abs}{\scriptsize absorption}
\psfrag{4.70636}{a)}
\includegraphics[width=15cm]{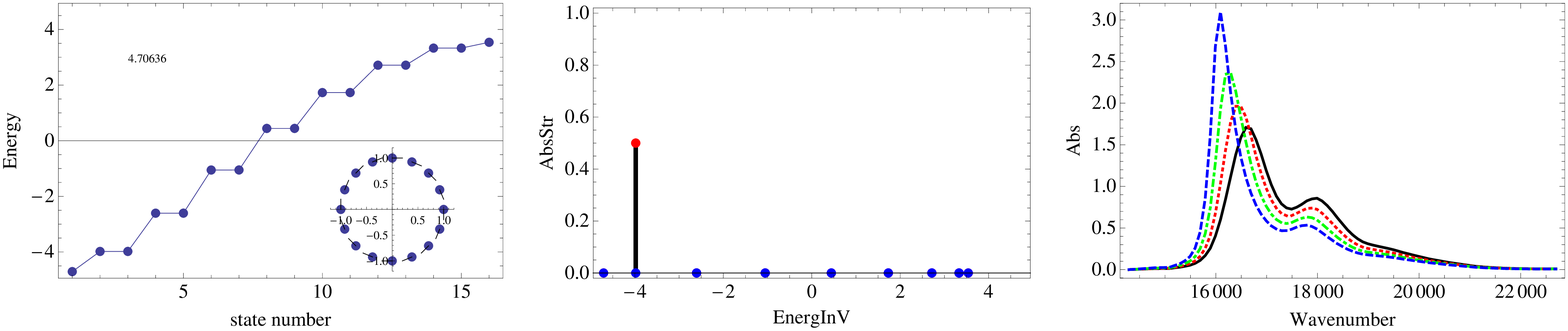}%
\vspace{-0.35cm}\\
\psfrag{13.5866}{b)}
\psfrag{state}{\scriptsize state number }
\psfrag{EnergInV}{energy}
\psfrag{Wavenumber}{\scriptsize energy in cm$^{-1}$}
\includegraphics[width=15cm]{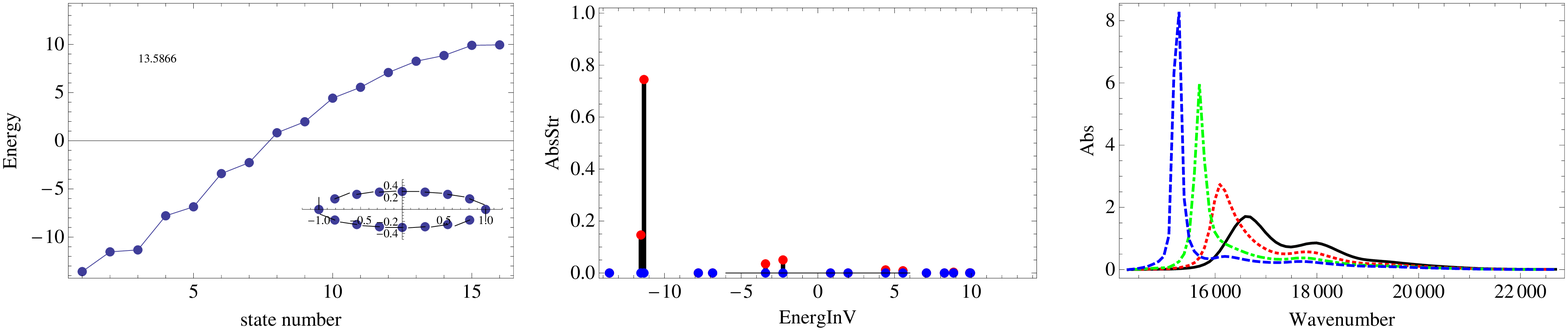}%
\caption{\label{fig:phi0}
As in Fig.~\ref{fig:Ring_RadTang} the dipoles are in the ring plane and $\phi=0^{\circ}$ wrt the tangent on the ring. The upper row shows $f=0$ (i.e.\ a ring) and the lower row a strong deformation $f=0.7$.
The CES spectra in the right column are for $V=50$, $100$ and $150$ cm$^{-1}$ (red, green , blue).
}
\end{figure*}

Having illustrated the extreme sensitivity of spectra to dipole orientation on a perfectly symmetric circular structure, we examine next the effect of symmetry-breaking by deforming the circle successively into an ellipse of increasing eccentricity. Rather than eccentricity, we will use the simpler 'flattening factor' $f= 1 - (b/a)$, where $b$ and $a$ are the minor and major axes of the ellipse, respectively. Hence, $f=0$ for a circle and $f=1$ when the circle is squashed flat. We consider two dipole arrangements on the circle, the first of which is relatively insensitive to deformation, the second of which shows strong sensitivity.\\
The first geometry is the tangential arrangement of Fig(\ref{fig:Ring_RadTang}a), shown again in Fig.(\ref{fig:phi0}a). In the right-hand column we now show the CES vibronic spectra. For a circle, the spectrum shows the characteristic J-band structure for strong coupling. In fact as the circle is squashed into an ellipse there is almost no perceptible change in the CES spectrum. Therefore, in Fig.~(\ref{fig:phi0}b) we show only the rather extreme case $f=0.7$, where the similarity to the circular case is confirmed. The reason for this can be explained by examination of the exciton wave-functions (not shown). As the ellipse is deformed,  states localised at the corners of the ellipse are formed but they remain in the middle of the exciton band and do not absorb appreciably. The two long segments of the ellipse become more and more like two linear chains. In this orientation, these have the same J-band absorption characteristics as the circle. The small changes that do occur are washed out in the CES spectrum, so that the right hand columns of Figs.(\ref{fig:phi0}a and \ref{fig:phi0}b) show spectra which are quite similar.\\

\begin{figure*}[tp]
\psfrag{state}{}
\psfrag{number}{}
\psfrag{EnergInV}{}
\psfrag{Wavenumber}{}
\psfrag{Energy}{\scriptsize Energy}
\psfrag{AbsStr}{\scriptsize absorption}
\psfrag{Abs}{\scriptsize absorption}
\psfrag{0.129454}{a)}
\includegraphics[width=15cm]{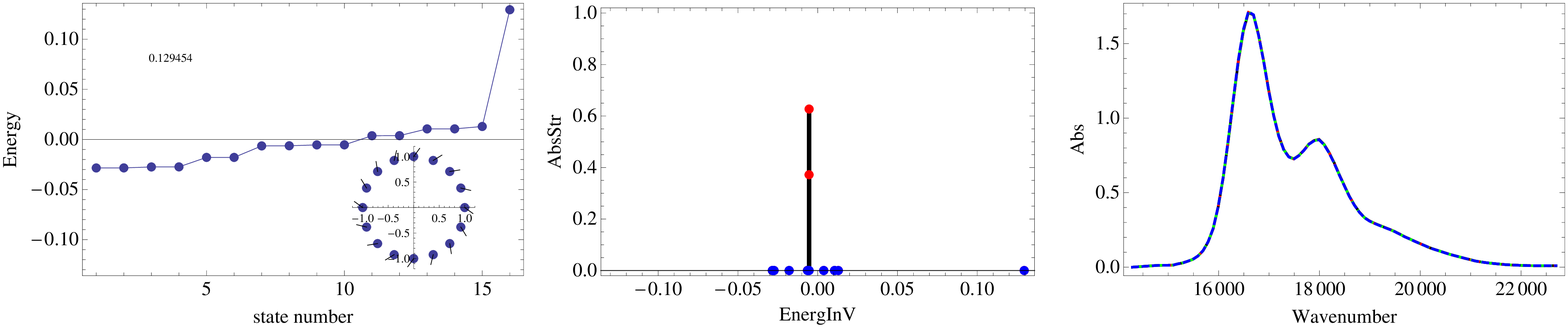}%
\vspace{-0.35cm}\\
\psfrag{0.254178}{b)}
\includegraphics[width=15cm]{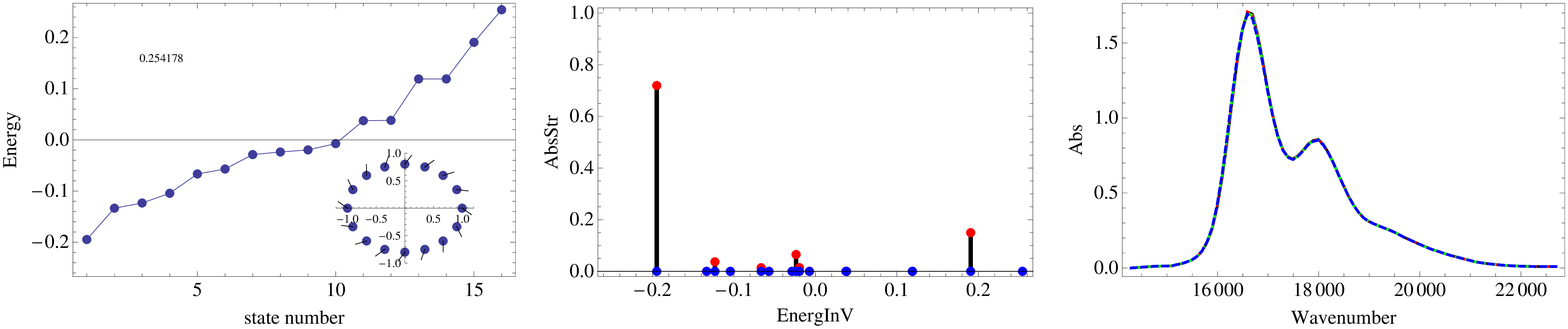}%
\vspace{-0.35cm}\\
\psfrag{0.697177}{c)}
\includegraphics[width=15cm]{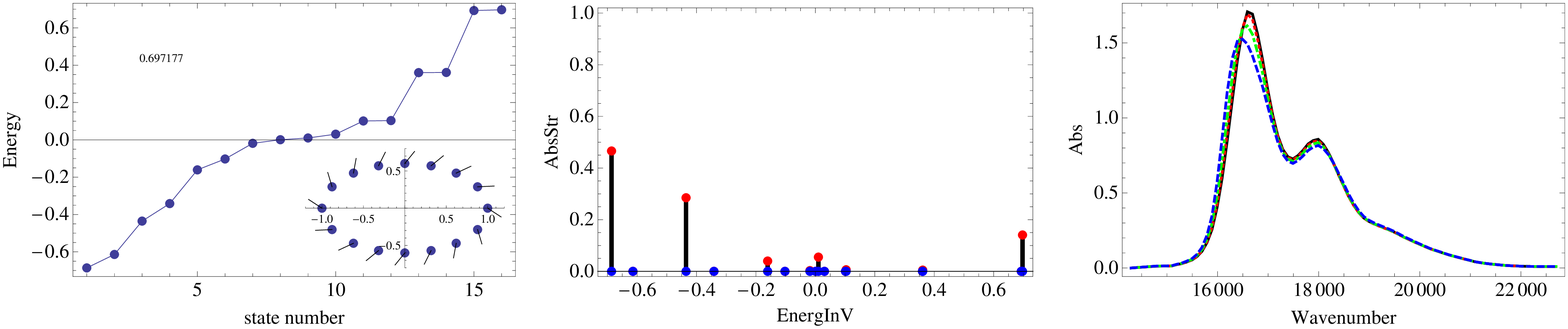}%
\vspace{-0.35cm}\\
\psfrag{4.46251}{d)}
\psfrag{state}{\scriptsize state number }
\psfrag{EnergInV}{energy}
\psfrag{Wavenumber}{\scriptsize energy in cm$^{-1}$}
\includegraphics[width=15cm]{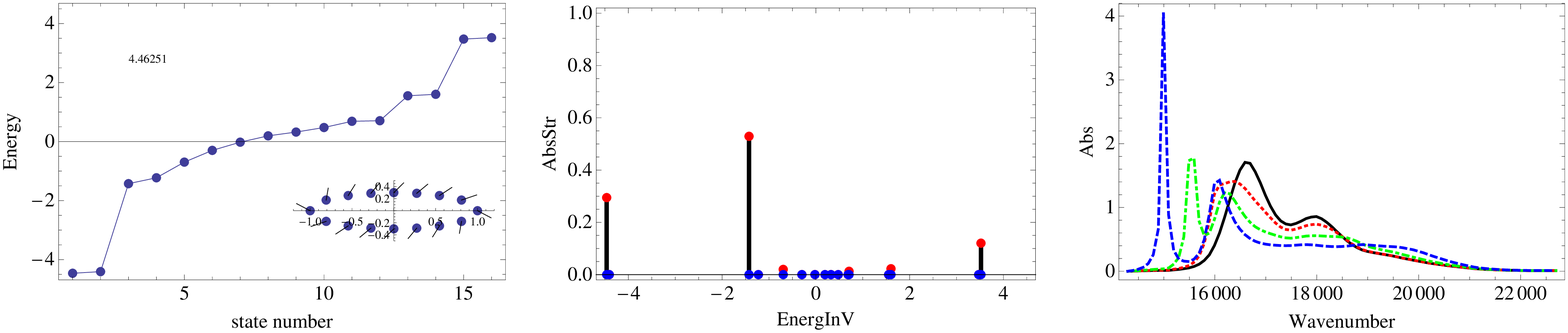}%
\caption{\label{fig:54}As in Fig.~(\ref{fig:phi0}) but for in-plane dipoles with orientation $\phi=54^{\circ}$ w.r.t. the tangent and aggregates with a) $f=0$, b) $f=0.2$, c) $f=0.4$, d) $f=0.7$.
The CES spectra in the right column are for $V=150$, $300$ and $450$ cm$^{-1}$ (red, green , blue).
}
\end{figure*}

As an arrangement which is extremely sensitive to symmetry-breaking we choose the case of in-plane dipoles with $\phi=54^\circ$ orientation of Fig.(\ref{fig:Ring_RadTang})d. This is reproduced as Fig.(\ref{fig:54}a). Here the nearest-neighbour coupling is essentially zero. One notes the two allowed levels sitting in the middle of a narrow band with the non-absorbing $j= 0$ level split off at the top of the band. The CES aggregate spectrum (right column) is then almost the same as the monomer spectrum.\\
 As the circle is flattened, spectacular changes occur as the nearest-neighbour coupling switches on. Already for $f=0.2$, Fig.(\ref{fig:54}b), the two degenerate allowed levels have split to be the bottom and, apart from $j=0$, top levels of the band. These states are still fully delocalised on the ellipse. However, as flattening progresses Figs.(\ref{fig:54}b,c), three new features arise. Firstly, the allowed levels acquire a degenerate partner and become the lowest and highest energy levels. Secondly, they become localised at the corners of the ellipse. Inspection of the wave-functions shows that, as in the linear case, they are nothing other than the $'+'$ and $'-'$ dimer levels, now doubly-degenerate and both dipole-allowed because of the orientation. The rest of the ellipse behaves as two linear segments so that the third feature to appear is a central exciton band with a new, fully delocalised J-band state at its lower edge and a weak H-band absorption at the upper edge. These features of dimer-pair formation and  the splitting off of localised states  to leave a central band, are seen clearly in the first and second columns of Fig.~(\ref{fig:54}d). The CES spectrum in the third column shows, in strong coupling, an apparent double J-band structure and a broad blue-shifted H band. However, we emphasise the completely different characters of the two apparently similar J bands. The lower is a localised state due to geometric disorder, the other is a delocalised exciton state. Comparison of the CES spectra of Fig.(\ref{fig:54}c) and Fig.(\ref{fig:phi0}b), both for $f=0.7$, emphasises the strong changes that can occur simply due to dipole orientation.\\

\begin{figure*}[tp]
\psfrag{state}{}
\psfrag{number}{}
\psfrag{EnergInV}{}
\psfrag{Wavenumber}{}
\psfrag{Energy}{\scriptsize Energy}
\psfrag{AbsStr}{\scriptsize absorption}
\psfrag{Abs}{\scriptsize absorption}
\psfrag{0.145995}{a)}
\includegraphics[width=15cm]{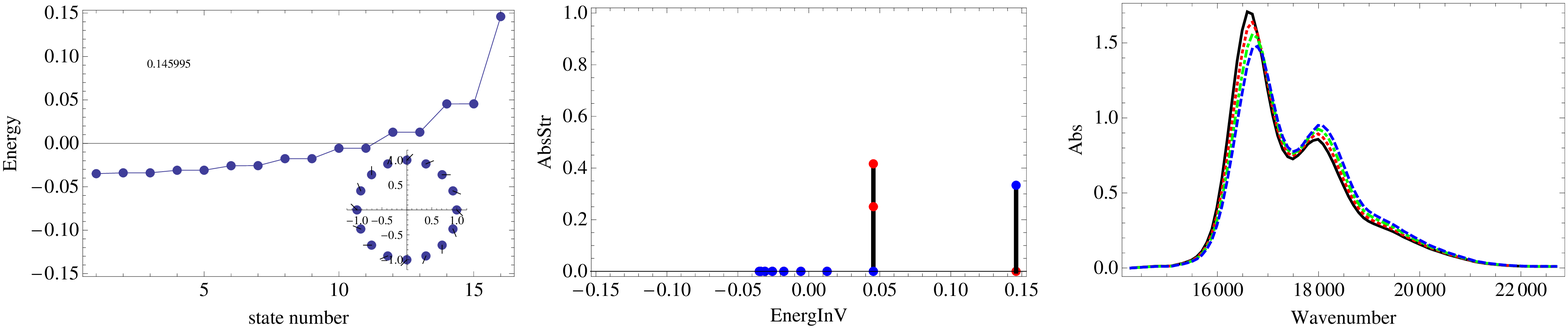}%
\vspace{-0.35cm}\\
\psfrag{0.248895}{b)}
\includegraphics[width=15cm]{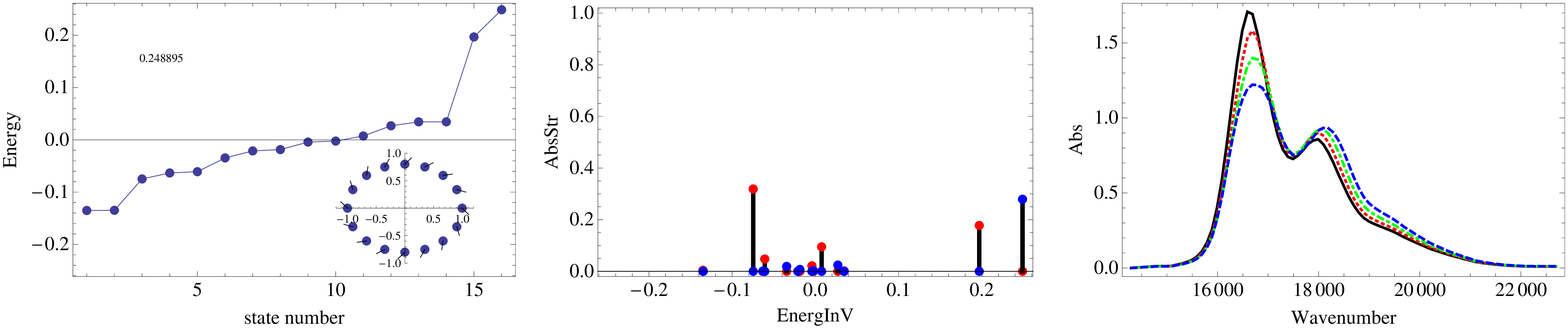}%
\vspace{-0.35cm}\\
\psfrag{0.575455}{c)}
\psfrag{state}{\scriptsize state number }
\psfrag{EnergInV}{energy}
\psfrag{Wavenumber}{\scriptsize energy in cm$^{-1}$}
\includegraphics[width=15cm]{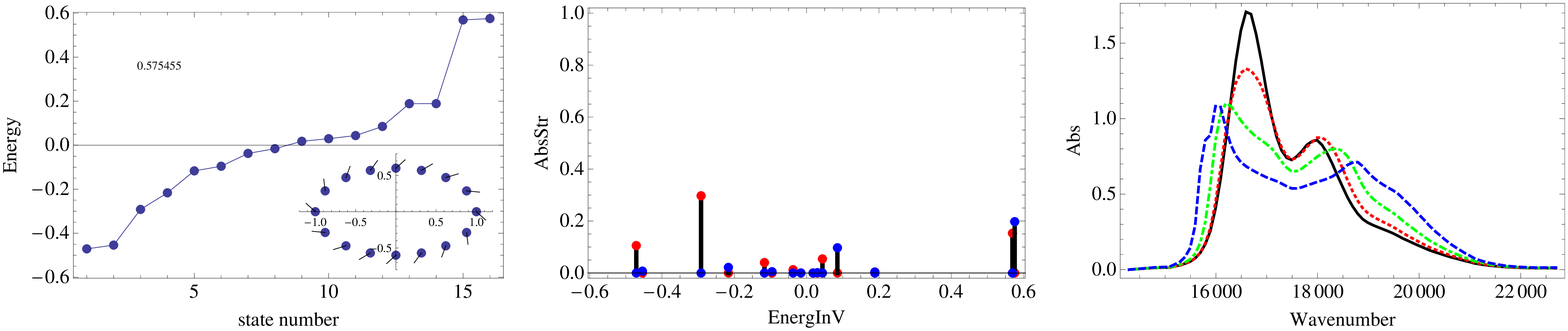}%
\caption{\label{fig:Ring_mu_all}As in Fig.~(6) but for  out-of-plane dipoles having $\phi = 45^\circ$ and polar angle $\theta = 55^\circ$ in the circular aggregate. The figures are for a) $f=0$, b) $f=0.2$, c) $f= 0.4$. The 
CES spectra are with  $V=750$  cm$^{-1}$, $1500$ cm$^{-1}$ and $2250$ cm$^{-1}$ (red, green , blue). 
}
\end{figure*}

  As a last example we consider out-of-plane orientation of the dipoles, with a $z$-axis perpendicular to the elliptical plane. We take all three $(x,y,z)$ dipole components to be equal corresponding to $\phi= 45^\circ$ and polar angle $\theta = 55^\circ$. From Fig.~(\ref{fig:Ring_mu_all}a) one sees that the circular geometry energy-level distribution (first column) is similar to the $\phi = 54^\circ$ case of Fig.~(\ref{fig:54}a), but the stick spectrum of the second column is quite different. Now the two $j=1$ absorbing levels with polarisation in the plane appear at the high-energy H-band side. In addition, the fully-symmetric, fully-delocalised $j=0$ level now carries oscillator strength polarised perpendicular to the plane. 

 Upon flattening to an ellipse, the changes mirror those of Fig.(\ref{fig:54}), except that the $j=0$ level maintains its dipole-allowed absorption throughout, since there are no changes in geometry in the perpendicular direction.  For $f=0.4$ the $j=1$ levels have acquired degenerate partners and separated, as states localised at the ellipse edges, from the bottom and top of the band of remaining levels. Of these levels, the lowest is again a delocalised exciton state carrying most of the oscillator strength and the highest levels now carry oscillator strength mostly polarised in the perpendicular direction. The result for the strong-coupling vibronic spectrum Fig.(\ref{fig:Ring_mu_all}c, right column) shows again a  J-band structure but now with a very pronounced blue-shifted H band apparent.
 
\section{Relevance for Experiments}

This study was motivated by observations of strong spectral changes accompanying pressure change giving rise to a  presumed flattening of cylindrical aggregates composed of different substituents of $5,5^\prime,6,6^\prime$ - tetrachlorobenzimidacarbocyanine (TDBC) dyes \cite{Sp99__,SpDa98_738_}. The following characteristics under increasing pressure were noted;

a) the non-aggregated monomer Stokes' shift increases linearly with pressure.

b) Unlike other studies on cyanine dyes \cite{LiCh96_5359_} and porphyrins \cite{ChHa97_9297_}
, the red shift of the J-band of the aggregate is not monotonically linear due to compression of the inter-monomer distance. Rather a linear shift is followed by saturation as pressure increases.

c) Nevertheless the aggregate Stokes' shift does increase monotonically with pressure.

d) As pressure increases a new absorption band arises as a shoulder on the spectrum at  the low-energy side of the J-band. 

Spitz and Daehne \cite{SpDa98_738_}  suggested that these changes arise from a collapse of the cylinder at a given pressure and formation of a localised state at the 'kink' in the now almost elliptical cross-section. The localisation of excitation on just two monomers implies that the Stokes' shift behaves in the same way as for the monomers (J-band emission shows a small Stokes' shift) and also can account for the new band emerging below the J-band. Unfortunately, due to the lack of knowledge of the precise geometry and particularly dipole orientation in such aggregates, we do not consider it realistic to present our detailed calculations showing these characteristics. Suffice it to say that the calculated spectra of elliptical conformations such as shown in Fig.~(\ref{fig:54}c and d) where a dipole-allowed localised state arises below the J-band level do lend qualitative support to the interpretation of Spitz and Daehne.\\

\section{Conclusions}
We have examined the influence of dipolar orientation and symmetry-breaking on the energy and light absorption spectra of  very simple linear and circular aggregates of identical monomers. The major features to appear are a strong dependence of absorption characteristics on orientation and the isolation of states localised around bends in regular structures. In many cases these localised states are split off from the top and bottom of the exciton band and may carry significant oscillator strength. Clearly the presence of such states, either dark or absorbing, below the exciton band edge can seriously affect the fluorescence properties and Stokes shift of the aggregate. The flattening of a circle also leads to the formation of localised states around the ellipse extremities. Depending upon dipole orientation these can lead to considerable broadening and blue shift of the spectrum. 
The results shown are admittedly model calculations, although we feel that the inclusion of vibrations in the CES approximation is realistic since in previous work we have obtained very good agreement with experiment. The aim of this study has been to draw attention to the localisation occurring due to very simple geometrical deformation of symmetric aggregates and to point out that this localisation can lead to new spectral structures.
\\

\end{document}